# Adaptive Ising machine based on phase-locking of an auto-oscillator to a bi-harmonic external driving with noise

Eleonora Raimondo[1], Andrea Grimaldi[2], Vasyl Tyberkevych[3], Riccardo Tomasello[2], Anna Giordano[4], Mario Carpentieri[2], Andrei Slavin[3], Massimo Chiappini[1], Giovanni Finocchio[1,5*]

[1]*Istituto Nazionale di Geofisica e Vulcanologia, Rome, 00143, Italy*
[2]*Department of Electrical and Information Engineering, Politecnico di Bari, Bari, 70126, Italy*
[3]*Department of Physics, Oakland University, Rochester, Michigan, 48309, USA*
[4]*Department of Engineering, University of Messina, I-98166 Messina, Italy*
[5]*Department of Mathematical and Computer Sciences, Physical Sciences and Earth Sciences, University of Messina, Messina, 98166, Italy*

[*]Corresponding author: gfinocchio@unime.it

**Abstract**

An auto-oscillator driven by a harmonic signal at about twice its free-running frequency is characterized by a bistable phase dynamics where the two states are separated by $\pi$ radians. This phase bistability enables an oscillator to emulate a single Ising spin, providing a fundamental building block for the oscillator-based Ising machines (OIM). At the same time, a driving signal close to the oscillator free-running frequency locks the oscillator's phase at a single value, playing the role of a magnetic field bias in ensembles of real spins.

We introduce a universal theory of phase auto-oscillators driven by a bi-harmonic signal (having frequency components close to single and double of the free-running oscillator frequency) with noise; with it, we show how deterministic phase locking and stochastic phase slips can be continuously tuned by varying the relative amplitudes and frequencies of the driving components. Using, as an example, a spin-torque nano-oscillator, we numerically validate this theory by implementing a deterministic Ising machine paradigm, a probabilistic one, and dual-mode operation of the two. This demonstration introduces the concept of adaptive Ising machines (AIM), a unified oscillator-based architecture that dynamically combines both regimes within the same hardware platform by properly tuning the amplitudes of the bi-harmonic driving relative to the noise strength.

Benchmarking on different classes of combinatorial optimization problems, the AIM exhibits complementary performance compared to OIMs and probabilistic Ising machines, with adaptability to the specific problem class. This work introduces the first OIM capable of transitioning between deterministic and probabilistic computation taking advantage of a proper design of the trade-off between the strength of phase-locking of an auto-oscillator to a bi-harmonic external driving and noise, opening a path toward scalable, CMOS-compatible hardware for hybrid optimization and inference.

Synchronization of nonlinear oscillators is a universal phenomenon that describes collective dynamics in physics, biology, and engineering. The Kuramoto synchronization model [1] is one of the most widely used theoretical frameworks for understanding the phase dynamics of externally driven auto-oscillators. In recent years, the phase dynamics described by [1] has emerged as a powerful framework for unconventional computing, enabling the mapping of oscillator phase states onto Ising spins for the solution of combinatorial optimization problems (COPs) [2–6]. Oscillator-based Ising machines (OIMs), which exploit second-harmonic injection locking to discretize oscillator phases, have demonstrated efficient deterministic minimization of the Ising Hamiltonian in optical, electronic, and spintronic systems [7–12]. A complementary approach is based on probabilistic Ising machines (PIMs), which leverages Gibbs sampling to generate a discrete dynamics that samples from the Boltzmann distribution, has been also developed recently [13–15]. While OIMs are effective at rapid deterministic convergence with a speed defined by their oscillation frequency, they lack the ability to explore solution spaces statistically. On the other hand, PIMs provide an effective platform for sampling but with a convergence speed fixed by the random number generation and the sequential Gibbs sampling protocol. To date, no oscillator-based system has provided both deterministic and probabilistic operation within a single architecture.

In this letter, we introduce the concept of adaptive Ising machine (AIM), an oscillator-based architecture that unifies the working principle of OIMs and PIMs on the same hardware platform. The AIM operates by dynamically tuning the balance between deterministic phase locking and stochastic phase fluctuations through continuous control of the oscillator's input parameters. Our AIM is different from the self-adaptive Ising machine proposed by Delacour [16], which archives adaptation at the algorithm level to handle constrained optimization problem. Building on a universal model of the phase dynamics in an auto-oscillator under composite and noisy driving, we developed an analytical theory describing how deterministic phase locking and stochastic phase slips can be continuously tuned by external control parameters. Using an example of a spin-torque nano-oscillator (STNO), we identify the potential limitations of OIMs for facing COPs having the local bias that is a phase difference between the two state phases of the injection locked oscillator different than π. We also demonstrate that AIMs can work as OIMs, as PIMs, and in a dual-mode operation regime, where robust deterministic binarization and tunable probabilistic sampling are key elements that provide a distinct form of hybridization at the device level. By integrating deterministic dynamics and probabilistic sampling within the same physical architecture – rather than combining heterogeneous systems – AIM can improve both feasibility and scalability compared to previous approaches that, for instance, combine classical Ising machines with post-processing, integrate annealers with digital solvers [17], apply constraint correction [18], decompose Ising problems into smaller sub-problems, or rely on coupling-based methods. We further benchmark AIM on instances of two COPs, maximum satisfiability problem (Max-SAT) and maximum cut problem (Max-Cut). AIM, therefore, establishes the first oscillator-based Ising machine capable of transitioning between deterministic and probabilistic computation, providing a scalable, CMOS-compatible approach with potential applications in hybrid combinatorial optimization and inference.

The dynamics of an auto-oscillatory system can be described by its phase $\varphi(t)$. Under weak periodic driving signal and noise, the phase dynamics in a general case is governed by the equation:

$$\frac{d\varphi}{dt} = \omega_0 - h(\omega_s t, \varphi) + \xi(t), \tag{1}$$

where $\omega_0$ is free-running frequency of the auto-oscillator, $h(\omega_s t, \varphi)$ accounts for the effect of the periodic driving signal with fundamental frequency $\omega_s$, and $\xi(t)$ is the thermal noise, modeled as white Gaussian noise with zero mean and second-order correlator $\langle \xi(t)\xi(t')\rangle = 2\Delta\omega_0 \delta(t-t')$, where $\Delta\omega_0$ is the half-linewidth of the free-running oscillator ($h = 0$). For slow phase dynamics, it is convenient to introduce the phase difference between the oscillator and the driving signal $\phi(t) = \varphi(t) - \omega_s t$. In this scenario, Eq. (1) can be written as $\frac{d\phi}{dt} = -\Omega_s - H(\phi) + \xi(t)$, where $\Omega_s = \omega_s - \omega_0$ is the frequency mismatch or detuning, and $H(\phi)$ is the averaged driving action. To describe the general case of an auto-oscillator under composite and noisy driving, we introduce the probability distribution function (PDF) $P_D(t, \phi)$ of the oscillator phase.

The standard stochastic analysis leads to the following Fokker-Planck equation:

$$\frac{\partial P_D}{\partial t} - \frac{\partial}{\partial \phi}\left[(\Omega_s + H(\phi))P_D\right] - \Delta\omega_0 \frac{\partial^2 P_D}{\partial \phi^2} = 0. \tag{2}$$

This equation describes the diffusion of an effective Brownian particle moving in a potential landscape defined by detuning, driving, and noise. For the driven oscillator ($\Omega_s + H(\phi) \neq 0$), the effective potential takes the form:

$$U(\phi) = \int_{\phi_0}^{\phi} [\Omega_s + H(\phi')] d\phi' = \Omega_s \phi - \sum_{n=1}^{\infty} \frac{\Lambda_n}{n} \cos[n(\phi - \beta_n)], \tag{3}$$

where the coefficient $\Lambda_n > 0$ is proportional to the amplitude of the $n$-th harmonics of the driving signal, and $\beta_n$ is the corresponding internal phase shift. The stationary solution of Eq. (2) is given by:

$$P_D(\phi) = p_0(1 + \rho G(\phi)) \exp\left(-\frac{U(\phi)}{\Delta\omega_0}\right), \tag{4}$$

with integration constant $p_0$ and corrections specified by $\rho$ and $G(\phi)$. Full derivations are provided in Sec. I of Supplemental Material [19].

The presence of a second harmonic driving term gives rise to two distinct stable phase states, $\phi_A > \pi$ and $\phi_B < \pi$, separated by $|\phi_A - \phi_B| = \pi$. In the absence of the first harmonic, these states are energetically equivalent, and the oscillator exhibits symmetric phase slips between them with equal probability. Adding a first harmonic driving breaks this symmetry, reshaping the energy landscape and biasing the system toward one of the two phases, as illustrated in Fig. 1(a). Detuning further modifies the dynamics by tilting the effective energy potential [20]. For negative detuning, the occupation probability of state $\phi_A$ as a function of $\Lambda_1$ exhibits a sigmoidal dependence, as shown in Fig. 1(b). In contrast, under the same condition but positive detuning, the potential becomes more complex, resulting in a non-monotonic dependence of the phase probability with $\Lambda_1$, as illustrated in Fig. 1(c). This non-monotonicity arises from the emergence of an additional minimum with $|\phi_A - \phi_B| < \pi$ (see Sec. II of Supplemental Material [19]) in the energy landscape.

Those results are valid for any auto-oscillatory system. Here we focus on STNOs realized with magnetic tunnel junctions (MTJs) and subjected to a bi-harmonics spin-transfer torque (STT) current and thermal noise. The STNO device considered has an elliptical cross-section, an out-of-plane free layer (FL) magnetization $\boldsymbol{m}$, and an in-plane polarizer magnetization, $\boldsymbol{m}_p$, as shown in the sketch in Fig. 1(d) [21]. This geometry allows for the realization of STNOs with ultralow threshold for the excitation of self-oscillations [22]. The FL dynamics is studied through micromagnetic simulations based on the Landau-Lifshitz-Gilbert-Slonczewski (LLGS) equation within the macrospin approximation [23,24]. The STNO is driven by an ac current with first- and second-harmonics, $J_{AC} = J_1 \sin(\omega_s t + \beta_1) + J_2 \sin(2\omega_s t + \beta_2)$, where $\omega_s = 2\pi f_s$ is the fundamental frequency. As a first step we analyze the effect of one component of the driving signal, $J_2 \sin(2\omega_s t + \beta_2)$. The STNO exhibits two synchronization regions (Arnold tongues) [20]: the first-harmonic injection locking (FHIL), where $2\omega_s$ is centered around the free-running frequency $\omega_0$, and the second-harmonic injection locking (SHIL), where $2\omega_s$ is centered around $2\omega_0$. Figure 1(d) shows the SHIL region, where two phase states separated by $\pi$ are stabilized. The black contour lines define the areas within the Arnold tongue where the injection-locking is stable for more than 90% and 80% of the simulation time. Details of the model and STNO characterization are provided in Sec. III of Supplemental Material [19]. The STNO locked dynamics is affected by the combined effect of the amplitudes of the bi-harmonic driving, by detuning effect that depends on the working point within the Arnold tongue, and by thermal noise strength.

At high $J_2$ (e.g., point C), the injection-locked oscillator remains fixed in one of the two stable phase states within the simulation time. As $J_2$ decreases, phase slip events are observed (e.g., point A), and their occurrence rate increases toward the boundary of the Arnold tongue. At lower values of $J_2$, the STNO's phase is no longer synchronized with the external signal. Thermal noise further enhances the occurrence of phase slips, promoting random transitions between the two stable phase states and reducing the overall locking stability [25–27]. As predicted analytically, the phases can be externally tuned by applying the first-harmonic driving component $J_1 \sin(\omega_s t + \beta_1)$, Figure 1(e) and (f) show the probability of the STNO being locked in phase $\phi_A$ as a function of the $J_1$, evaluated at two points chosen for their highest phase slips rates, within the region of negative (point A) and positive (point B) detuning, respectively. The behavior is in quantitative agreement with the theory. This observed trend is robust over a wide parameter range within the Arnold tongue (see Sec. IV of Supplemental Material [19]). Moreover, the locking probability can be further controlled by adjusting the internal phase $\beta_1$. Section V of Supplemental Material [19] summarizes the dependence of the locking probability of $\phi_A$ at point B as a function of $J_1$ for different values of $\beta_1$.

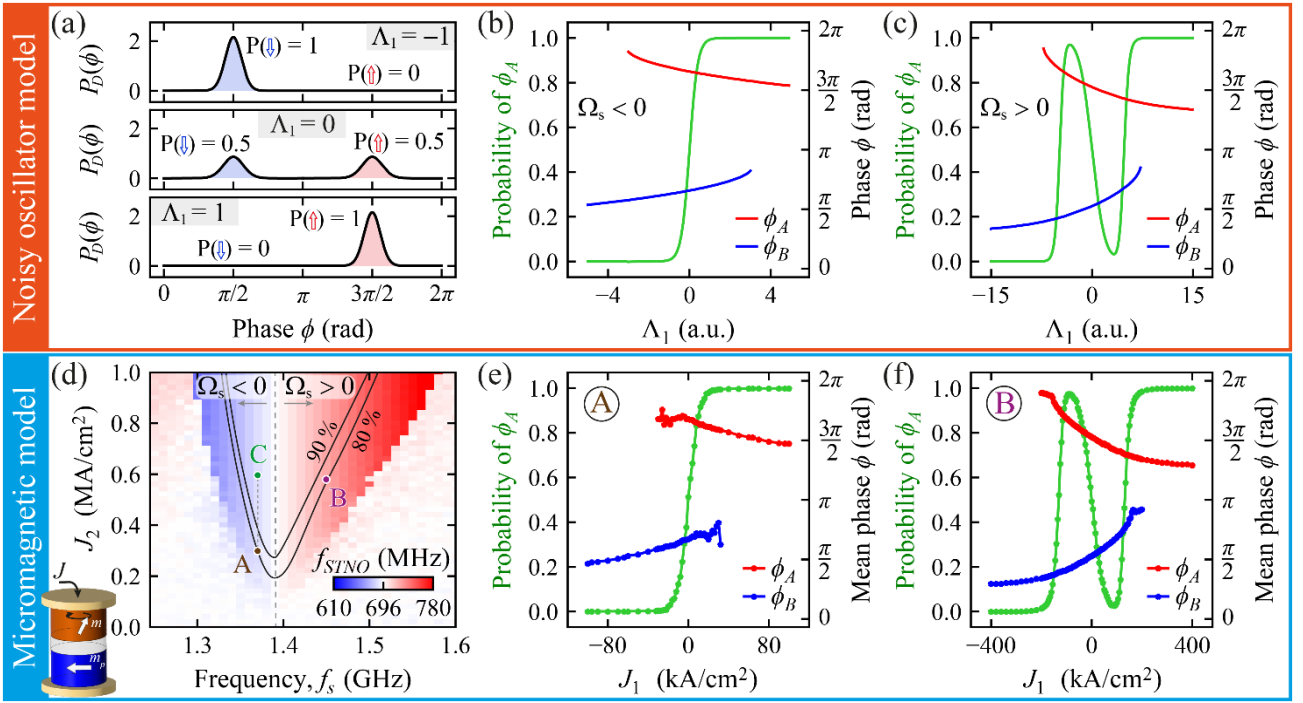

FIG. 1. A comparison of theoretical and numerical results for the oscillator's dynamics. (a) PDF of a noisy oscillator for different values of $\Lambda_1$. (b-c) Phase values from the analytical model for negative and positive detuning, respectively, and the probability dependence of $\phi_A$ on $\Lambda_2$ (green curve). (d) Arnold tongue region centered at $2f_0$. The point C is used for the simulations shown in Fig. 2. (e, f) Corresponding phase values and probability of $\phi_A$ from numerical simulations of STNO for negative (point A in panel (d)) and positive detuning (point B in panel (d)), respectively.

An STNOs operating in the SHIL regime with vortex-state static magnetization have already been shown to exploit phase slip dynamics to generate unbiased binary random numbers with 50% probability, proposing them as a building block for stochastic computing [25]. However, to realize PIMs based on STNOs, it is crucial that the probability of the phase slip events can be tuned via an external parameter with the necessary hyperbolic tangent (*tanh*) dependence [13]. Our findings demonstrate such a tunability, and highlight the importance of the working point position inside the Arnold tongue. Furthermore, by simply tuning $J_2$, the STNO can be driven from a deterministic regime with stable phase locking to a stochastic regime characterized by frequent phase slips. This fundamental property enables the design and the implementation of both OIMs and PIMs on a unified hardware platform, controlled by the strength of the relative amplitudes of the bi-harmonic driving signal. Building on this concept, we introduce the AIM, based on a network of *N* coupled STNOs, that collectively attempt to minimize the Ising Hamiltonian $H(\boldsymbol{\sigma}) = -\left(\frac{1}{2}\sum_{i,j} J_{ij}\sigma_i\sigma_j + \sum_i h_i\sigma_i\right)$, where $\sigma_i$ are the binary Ising spins, $J_{ij}$ is the coupling strength between the spins *i* and *j*, and $h_i$ is an external bias applied to the spin *i*. The time evolution of the normalized magnetization vector $\boldsymbol{m}_i$ of the FL in the $i^{th}$ oscillator $\forall i \in [1, N]$ is described by a system of *N* coupled LLGS equations. The coupling among the STNOs is implemented with the binarized phases $\sigma_i$, in analogy with the PIMs [13,14]. This approach overcomes the limitation of direct phase coupling in conventional OIMs. In this configuration, dedicated CMOS circuitry computes the effective coupling force applied to each oscillator according to $J_{i1} = \gamma\left(\sum_j J_{ij}\sigma_j + h_i\right)$, where $\gamma$ is the inverse pseudo-temperature used for implementing the annealing schedule, similarly to PIMs [13,15]. To evaluate the performance of the proposed AIM, we first assess the AIM working in the PIM regime at point A in Fig. 1(d). The phase probability in Fig. 1(e) follows a *tanh* dependence with slope of $10^{-4}$, which we use as a scaling constant for the inverse pseudo-temperature $\gamma$. Sampling capability is tested by implementing the invertible AND gate, mapped onto a three-node Ising model. The resulting phase-based probability distribution closely matches the expected Boltzmann distribution, confirming that the system correctly samples all four valid logic states with uniform probability (see Sec. VI of Supplemental Material [19]).

We next benchmark the AIM on two representative COPs, with approximately 100 Ising spins – corresponding to a network of 100 STNOs – and compare its performance against the OIM and the PIM. As discussed above, the STNO-based IM supports three modes: (i) OIM, stable deterministic injection-locking at large $J_2$ (point C in Fig. 1(d)); (ii) PIM, stochastic phase locking at small $J_2$ (point A in Fig. 1(d)); and (iii) AIM, where $J_2$ is linearly ramped from 0 to point C, enabling a continuous transition from unlocked to stochastic and, then, to deterministic dynamics. Figure 2(a-c) summarizes the results for the Max-Cut instance "gka1d" (101 nodes) from the BiqMac library, and Figure 2(d-f) show the performance on a Max-SAT instance with 20 clauses and 91 variables (112 spins). In both cases, we apply a standard simulated annealing schedule, with $\gamma$ linearly increased over a total simulation time of 75 $\mu s$. Success probabilities, calculated over 100 trials, are reported in the sub-figure titles, with average solution times in parentheses. The results reveal clear performance distinctions across the operating modes. The deterministic OIM struggles to explore the solution space. Both the PIM and the AIM achieve high success probabilities, with the PIM outperforming the AIM on the Max-Cut instance, whereas the AIM exhibits superior performance on the Max-SAT instance. These findings demonstrate that no single mode universally dominates, emphasizing the importance of adaptive strategies that dynamically select or combine computational regimes according to the problem.

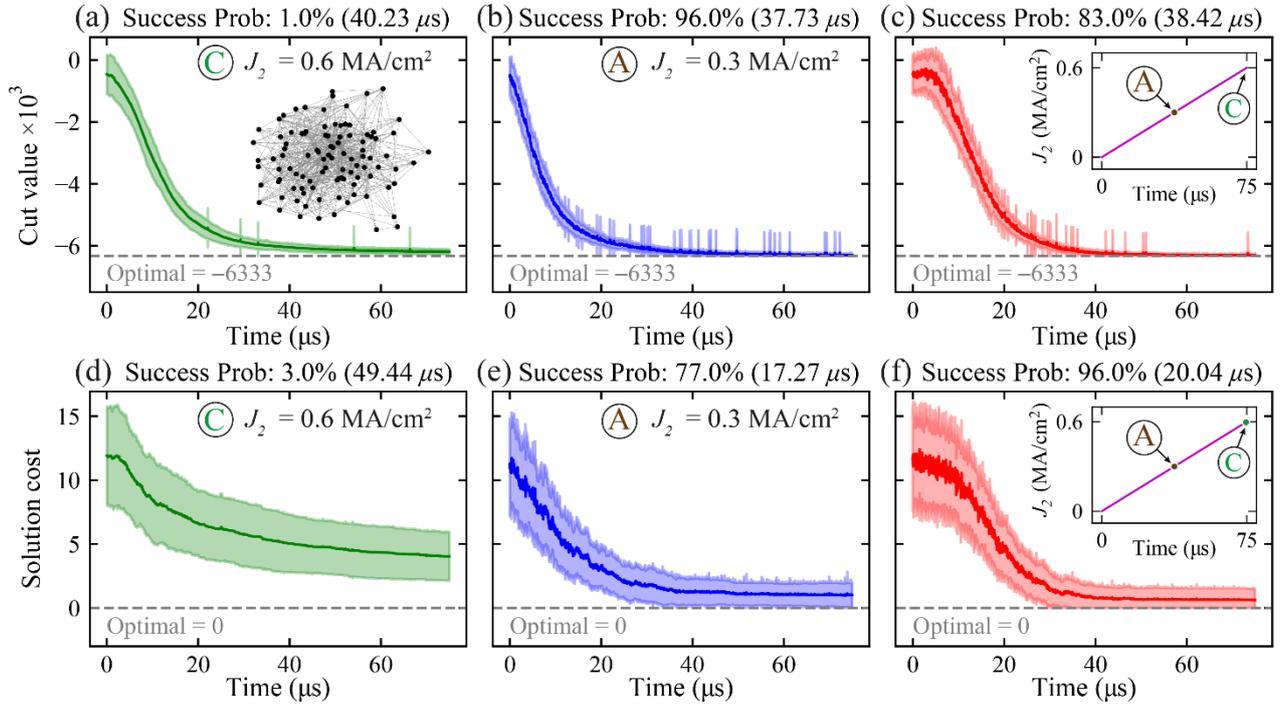

FIG. 2. Performance benchmark of the different types of STNO-based IMs by using simulated annealing as the energy minimization algorithm. (a-c) Cut value over time for the Max-Cut problem instance "gka1d" on a 101-node graph (inset in a), evaluated in three different operating modes: (a) OIM – point C ($J_2 = 0.6$ MA/cm²) (b) PIM – point A ($J_2 = 0.3$ MA/cm²) and (c) AIM: $J_2$ is linearly ramped from 0 to 0.6 MA/cm², enabling a dynamic transition from unlocked state to OIM. (d-f) Time evaluation of solution cost for a Max-SAT problem instance with 112 Ising spins, evaluated under the same three operating modes as in (a-c). The solid lines represent the average loss as a function of simulation time, with shaded areas indicating the first standard deviation of the distribution over 100 trials. The dashed gray lines mark the optimal value for each COPs.

We have introduced the idea of AIM, an oscillator-based architecture that unifies deterministic and probabilistic regimes within a single hardware platform, rather than combining heterogeneous approaches. By establishing a universal theory of auto-oscillators under bi-harmonics noisy driving, and validating it with STNOs, we demonstrated how AIMs exploit transitions between deterministic dynamics and tunable probabilistic sampling by tuning the phase slip probability in facing different classes of COPs. The tunability of the AIM working regime is directly linked to the strategy for developing Ising solvers that implement hybrid approaches, either digital-analog or quantum-classical, which are expected to be essential for overcoming the scaling limits of any single paradigm [28,29]. These features not only broaden the scope of problems that OIMs can address, but also point toward scalable, CMOS-compatible hardware for hybrid optimization and sampling. Notably, spintronic technology and processes for co-integrating MTJs with CMOS are already well-established in several fabs [30]. In this context, AIM complements algorithmic hybrid approaches, and positions oscillator-based systems as promising candidates for the next-generation of Ising machines.

*Acknowledgments* – This work has been supported by project number 101070287 — SWAN-on-chip — HORIZON-CL4-2021-DIGITAL-EMERGING-01, the projects PRIN 2020LWPKH7 "The Italian factory of micromagnetic modeling and spintronics", PRIN20222N9A73 "SKYrmion-based magnetic tunnel junction to design a temperature SENSor—SkySens", PON Capitale Umano (CIR_00030), D.D. n. 47 del 20 febbraio 2025 - assunzione di ricercatori internazionali post-dottorato: CUP: D53C25000620007, and by the PETASPIN association ([www.petaspin.com](www.petaspin.com)). R.T. and M.C. acknowledge support from the Project PE0000021, "Network 4 Energy Sustainable Transition – NEST", funded by the European Union – NextGenerationEU, under the National Recovery and Resilience Plan (NRRP), Mission 4 Component 2 Investment 1.3 - Call for tender No. 1561 of 11.10.2022 of Ministero dell'Università e della Ricerca (MUR) (CUP C93C22005230007).

*Data availability* – The data are available from the authors upon reasonable request.